\documentclass{aa}  

\usepackage{graphicx}
\usepackage{txfonts}
\usepackage{xcolor}
\usepackage{soul}
\usepackage{subcaption}  % for subfigure environment
\usepackage{graphicx}
\begin{document}

   \title{Understanding coronal rain dynamics through a point-mass model}

%   \subtitle{}

   \author{Andrew Hillier\inst{1} \and Ramon Oliver\inst{2,3} 
          \and
          David Mart\'{i}nez-G\'{o}mez\inst{4,5}
          }

   \institute{Department of Mathematics and Statistics, University of Exeter, Exeter, EX4 4QF UK\\
              \email{a.s.hillier@exeter.ac.uk}
         \and
             Departament de F\'{i}sica, Universitat de les Illes Balears, E-07122, Palma de Mallorca, Spain
         \and
              Institut d'Aplicacions Computacionals de Codi Comunitari (IAC3), Universitat de les Illes Balears, E-07122, Palma de Mallorca, Spain
         \and
             Instituto de Astrof\'{i}sica de Canarias, E-38205 La Laguna, Tenerife, Spain
         \and
              Departamento de Astrof\'{i}sica, Universidad de La Laguna, E-38205 La Laguna, Tenerife, Spain
             }

   \date{}

% \abstract{}{}{}{}{} 
% 5 {} token are mandatory
 
  \abstract
  % context heading (optional)
  % {} leave it empty if necessary  
   {}
  % aims heading (mandatory)
   {Observations and simulations of coronal rain show that as cold and dense plasma falls through the corona it initially undergoes acceleration by gravity before the downward velocity {saturates. Simulations have shown the emergence of an unexpected relation between terminal velocity of the rain and density ratio that has not been explained. Our aim is to explain this relation.}}
  % methods heading (mandatory)
   {In this paper we develop a simple point-mass model to understand how the evolution of the ambient corona moving with the coronal rain drop can influence the falling motion.}
  % results heading (mandatory)
   {We find that this simple effect results in the downward speed reaching a maximal value before decreasing, which is consistent with simulations with realistic coronal rain mass. These results provide an explanation for the scaling of the maximum downward speed to density ratio of the rain {to the corona and} as such provide a new tool that may be used to interpret observations.}
  % conclusions heading (optional), leave it empty if necessary 
   {}


   \maketitle
%
%-------------------------------------------------------------------

\section{Introduction}

Quiescent coronal rain is a {multiphase plasma phenomenon of cool plasma embedded in a hot component} that is frequently observed in solar active regions \citep{antolin2020, antolin2022, sahin2022,sahin2023ApJ...950..171S}. It starts with the continuous evaporation of cold and dense plasma from the solar chromosphere into the corona. This evaporated plasma, that now has the typical hot and rarefied conditions of the corona, accumulates {at the apex of} a coronal loop and eventually becomes thermally unstable {\citep{Parker1953ApJ...117..431P,Field1965ApJ...142..531F,Antiochos1980ApJ...241..385A}}, which leads to the formation of a dense coronal rain clump some tens of Mm high in the corona. Electron {number} densities of the order of {$2 - 25\times 10^{10}$~cm$^{-3}$} have been measured by \citet{antolin2015}, {compared to characteristic coronal loops densities in active regions of the order of $1 - 10 \times 10^{9} \ \rm{cm^{-3}}$ \citep{Reale2010}}. 
{These values translate  into clump densities of the order of 3–40 $\times 10^{-14}$~g~cm$^{-3}$ \citep[note that post-flare coronal rain can reach higher densities; see][who report values $\sim 10^{-12}$~g~cm$^{-3}$]{sahin2024}.} 
%\st{\mbox{\citep[somewhat smaller values of $1.4-2.3\times 10^{-14}$~g~cm$^{-3}$ have been given by][]{sahin2024}}} \st{compared to characteristic coronal densities (at a height of $\sim 0.1$ solar radius above the surface) of $n_{\rm{e}} \approx 10^{8 - 8.6} \ \rm{cm^{-3}}$} \st{\mbox{\citep{Dudik2021ApJ...906..118D, DelZanna2023ApJS..265...11D}}}. 
After this formation phase, coronal rain falls along the loop legs toward the solar surface under the action of gravity, although {in many cases it does not attain} the gravitational free-fall speed, but instead reaches a constant {(or even decreasing)} velocity. Coronal rain typically reaches falling speeds of up to 150~km~s$^{-1}$ \citep[see, for example,][]{Wiik1996SoPh..166...89W,antolin2012,kriginsky2021}.

{The descending phase of coronal rain is characterised by the cold and dense plasma flowing along coronal loops, which are aligned with the coronal magnetic field {\citep[see, e.g.,][]{Tripathi2006A&A...449..369T,Tripathi2007A&A...472..633T}} implying that some inherent aspects of the dynamics are essentially one-dimensional (1D).} This was used by \citet{Oliver2014} to set up 1D hydrodynamic models of coronal rain that reproduce the observed smaller-than-free-fall speed. These authors also found that coronal rain blobs achieve a maximum descending velocity, after which their speed is reduced gradually in time. The work of \citet{Oliver2014} was extended to 2D by \citet{Martinez2020}. A very interesting aspect of the simulations of \citet{Martinez2020} is that in the limit of strong magnetic field, i.e., $|B|\gtrsim 20$~G, the simulated coronal rain dynamics effectively became the sum of a set of vertical 1D models of the type presented in \citet{Oliver2014}, meaning that 1D modelling is reasonable given the strong magnetic field of the solar corona {as this implies the Alfv\'{e}n Mach number and plasma beta of the coronal rain is small} \citep[magnetic field strengths of the order of hundreds of Gauss or even larger have been inferred by][]{schad2016, kuridze2019, kriginsky2021}. {Both in \citet{Martinez2020} and in other multi-dimensional simulations using weaker field strengths  \citep[e.g.][]{Fang_2015} it has been found that there is a greater influence of multi-dimensional effects.}
Assessing the dynamics of their simulated rain blobs, \citet{Martinez2020} found, through empirical fit, that the maximum downward velocity of the falling coronal rain scales as the {rain to corona} density ratio {to} the power $0.64$ {(i.e. $(\rho_{\rm rain}/\rho_{\rm corona})^{0.64}$)}. A key question that still remains is: why this particular exponent?

One possible way of understanding the dynamics of these simulations may be through simplified point-mass and drag models. These models have been widely applied in many areas of fluid dynamics including the modelling of coronal mass ejection (CME) propagation \citep[e.g.][]{Vrsnak2010, Vrsnak2013}. A standard aspect of drag models is that the faster the body moves, the stronger the drag force works to decelerate it. Physically, this relates to the moving structure having to do more work to push aside the ambient material as it moves faster. 
This model predicts a terminal fall velocity that scales (in the large density contrast limit) as approximately the density ratio to the power $0.5$ \citep[e.g.][]{Zhou2021}. This discrepancy to the exponent found by \citet{Martinez2020} for their simulations is interesting as it implies that some different physics needs to be considered.   

Another key aspect of point-mass models, including drag models, is that as the mass moves through the medium, a portion of the ambient fluid, which will be referred to as the virtual mass, moves with the body. In effect, this adds to the mass of the moving body. For a standard drag problem, including one that involves incompressible flow, the virtual mass of the system is expected to not change significantly over time \citep[e.g. as used in ][]{Vrsnak2010, Vrsnak2013}. %\footnote{Maybe we can emphasise that in our system the virtual mass does change in time. Btw, we have not plotted $m_v(t)$: this would allow us to see if the virtual mass becomes comparable to $m_b$ and, if so, when this happens and to link it with the evolution of $v_b(t)$.} 
If the mass of the body is larger than that of the virtual mass, then the virtual mass does not change the dynamics in any significant way. Due to the high mass of a coronal rain blob, we may naturally expect this to be the case. However, this assumption may not hold for the coronal rain falling along magnetic field as a significant amount of material in front of and behind the moving coronal rain has to move as well. 

In this paper we develop a simple model for the coronal rain dynamics simulated by \citet{Oliver2014} and \citet{Martinez2020} looking at the role of virtual mass to quantify how it controls the evolution of the rain motion. We  then use this model to provide an explanation for the density and coronal rain velocity scaling presented in \citet{Martinez2020}.

\section{An overview of the numerical modelling}\label{model}

We consider a one-dimensional vertical slice of a fully ionised hydrogen plasma. The adiabatic hydrodynamic evolution of this system is described by the continuity equations of mass, momentum and {pressure}:

\begin{align}
    \frac{\partial\rho}{\partial t} &= -v\frac{\partial\rho}{\partial z} -\rho\frac{\partial v}{\partial z}, \label{eq:mass} \\
    \rho\frac{\partial v}{\partial t} &= -\rho v\frac{\partial v}{\partial z} -\frac{\partial p}{\partial z} -\rho g, \label{eq:momentum} \\
    \frac{\partial p}{\partial t} &= -v\frac{\partial p}{\partial z} -\gamma p\frac{\partial v}{\partial z}. \label{eq:energy}
\end{align}
Here $t$ and $z$ are time and the vertical coordinate, with the $z$-axis pointing upward; $\rho(z,t)$, $v(z,t)$ and $p(z,t)$ are the density, vertical velocity component and pressure; $g = 274$~m~s$^{-2}$ is the acceleration of gravity at the solar surface; and $\gamma=5/3$ is the ratio of specific heats.

In an isothermal atmosphere with temperature $T_0$ the plasma is in hydrostatic equilibrium with pressure and density given by

\begin{align}
    p(z) = p_0 \exp\left(-\frac{z}{H}\right), \label{eq:pnot} \\
    \rho(z) = \rho_0 \exp\left(-\frac{z}{H}\right), \label{eq:rhonot}
\end{align}

\noindent where the vertical scale-height $H$ is

\begin{equation}
    H = \frac{2 k_B T_0}{m_p g}, \label{eq:scale-height}
\end{equation}

\noindent with $k_B$ the Boltzmann constant and $m_p$ the proton mass. To derive Equation~\ref{eq:scale-height} we have used the perfect gas law for a fully ionised hydrogen gas,

\begin{equation}
    p_0 = 2\frac{k_B}{m_p}\rho_0 T_0. \label{eq:gas-law}
\end{equation}

Now, we study the temporal evolution of the system with zero initial velocity, initial pressure given by Equation~\ref{eq:pnot} and initial density given by the sum of Equation~\ref{eq:rhonot} plus a blob density,

\begin{equation}
    \rho_b(z) = \rho_{b0} \exp\left[-\left(\frac{z-z_0}{\Delta}\right)^2\right].
\end{equation}

\noindent To obtain $\rho(z,t)$, $v(z,t)$ and $p(z,t)$ we perform numerical simulations as described in \citet{Oliver2014}. The problem parameters are the background homogeneous temperature, $T_0$, the density at the $z=0$ height, $\rho_0$, the initial height of the dense blob, $z_0$, and the blob density and length, $\rho_{b0}$ and $\Delta$. The pressure at the $z=0$ height, $p_0$, appears in Equation~\ref{eq:pnot} and can be determined from $T_0$ and $\rho_0$ with the help of Equation~\ref{eq:gas-law}. With no loss of generality, we set $\rho_0=5\times 10^{-12}$~kg~m$^{-3}$ and $z_0=50$~Mm. Hence, we are left with three free parameters, namely $T_0$ (or $H$), $\rho_{b0}$ and $\Delta$.

\begin{figure*}[ht]
  \begin{center}
    \begin{subfigure}[t]{0.48\textwidth}
        \centering
        \includegraphics[width=\textwidth]{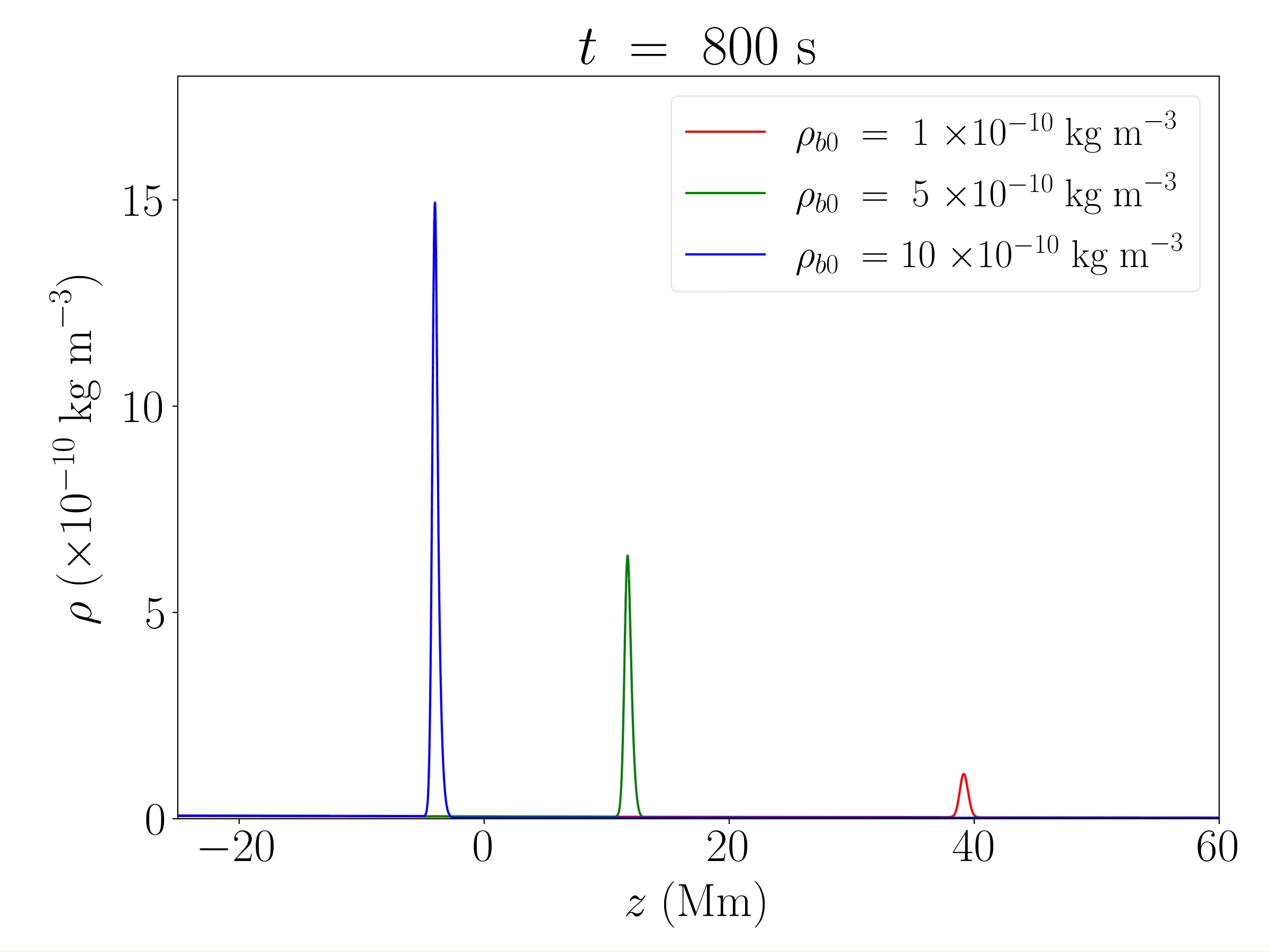}
        \put(-230, 180){\textbf{(a)}}
    \end{subfigure}
    \hfill
    \begin{subfigure}[t]{0.48\textwidth}
        \centering
        \includegraphics[width=\textwidth]{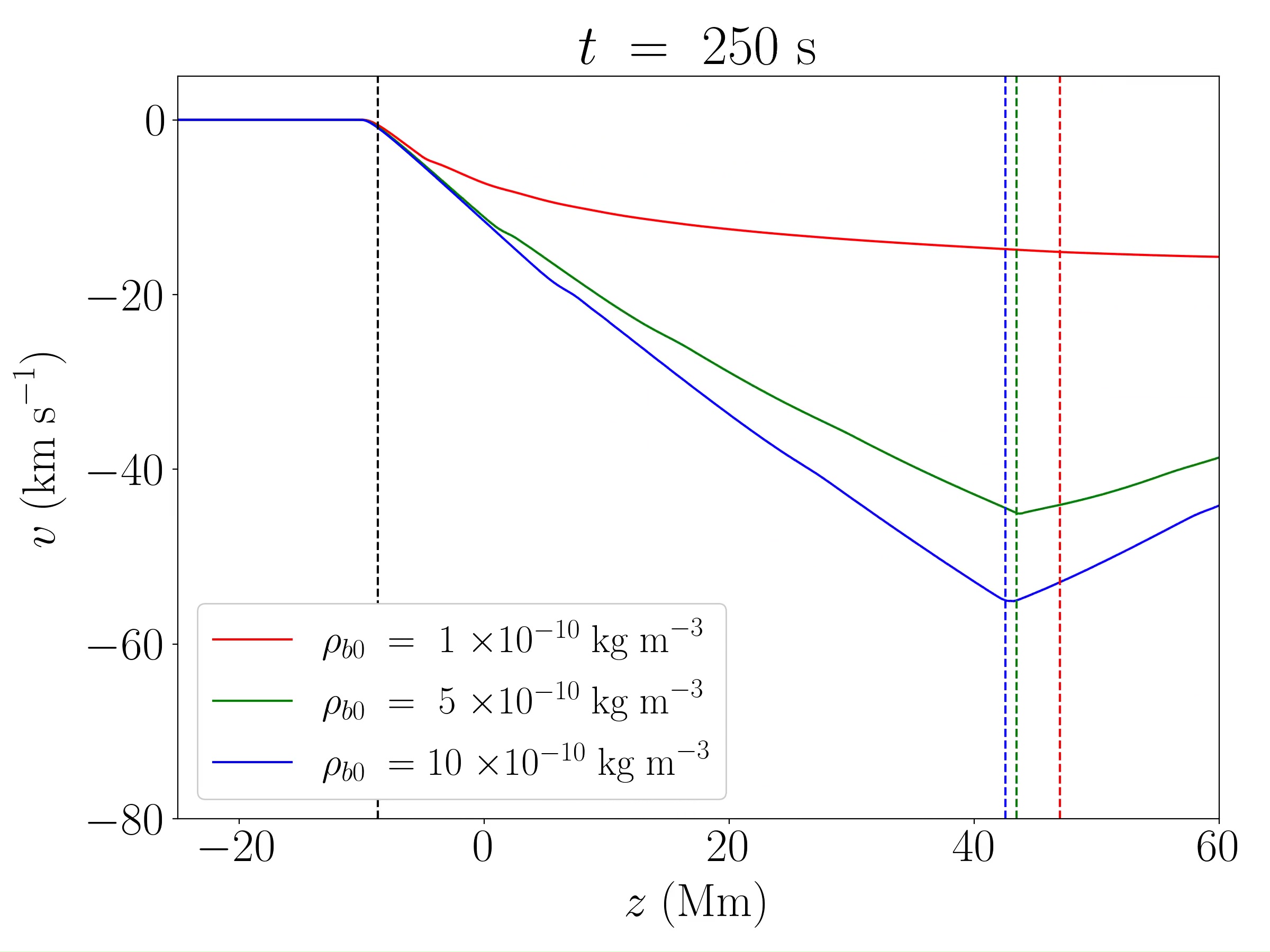}
        \put(-230, 180){\textbf{(b)}} 
    \end{subfigure}
  \end{center}
  \caption{Results of three numerical simulations with with $T_0=2\times 10^6$~K, $\Delta=0.5$~Mm and $\rho_{b0}=10^{-10}$~kg~m$^{-3}$ (red), $\rho_{b0}=5\times10^{-10}$~kg~m$^{-3}$ (green) and $\rho_{b0}=10\times10^{-10}$~kg~m$^{-3}$ (blue). (a) Density as a function of height at a given time. (b) Velocity as a function of height at a given time. A dashed black vertical line is plotted at $z=z_0-C_st$ {giving the position of the front of the compression wave}. The red, green and blue dashed vertical lines give the position of the maximum density. Animations of these figures can be found online.}%
\label{fig:simulation_results}
\end{figure*}

\begin{figure*}[ht]
  \begin{center}
    \begin{subfigure}[t]{0.48\textwidth}
        \centering
        \includegraphics[width=\textwidth]{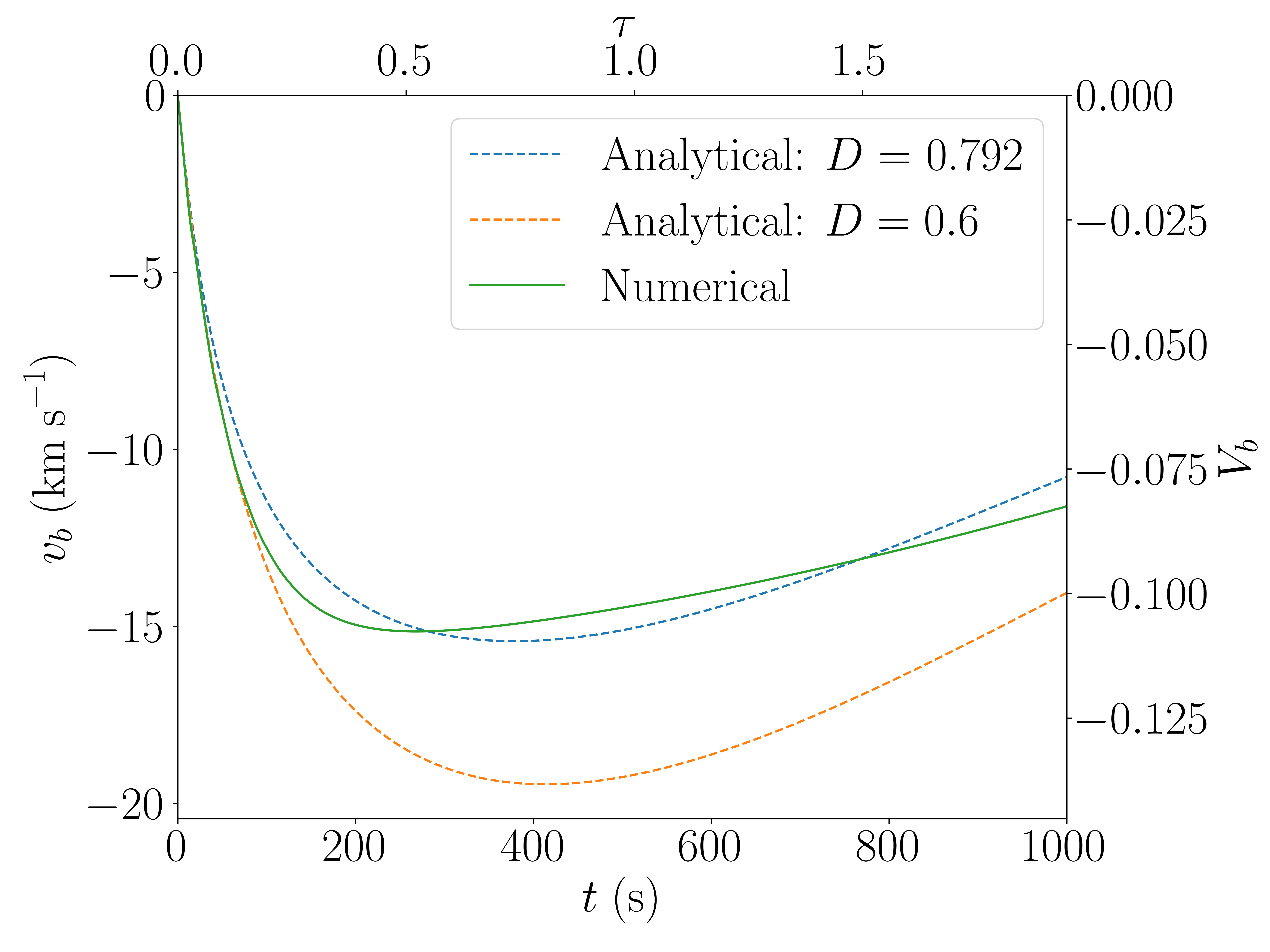}
        \put(-240, 180){\textbf{(a)}}
    \end{subfigure}
    \hfill
    \begin{subfigure}[t]{0.48\textwidth}
        \centering
        \includegraphics[width=\textwidth]{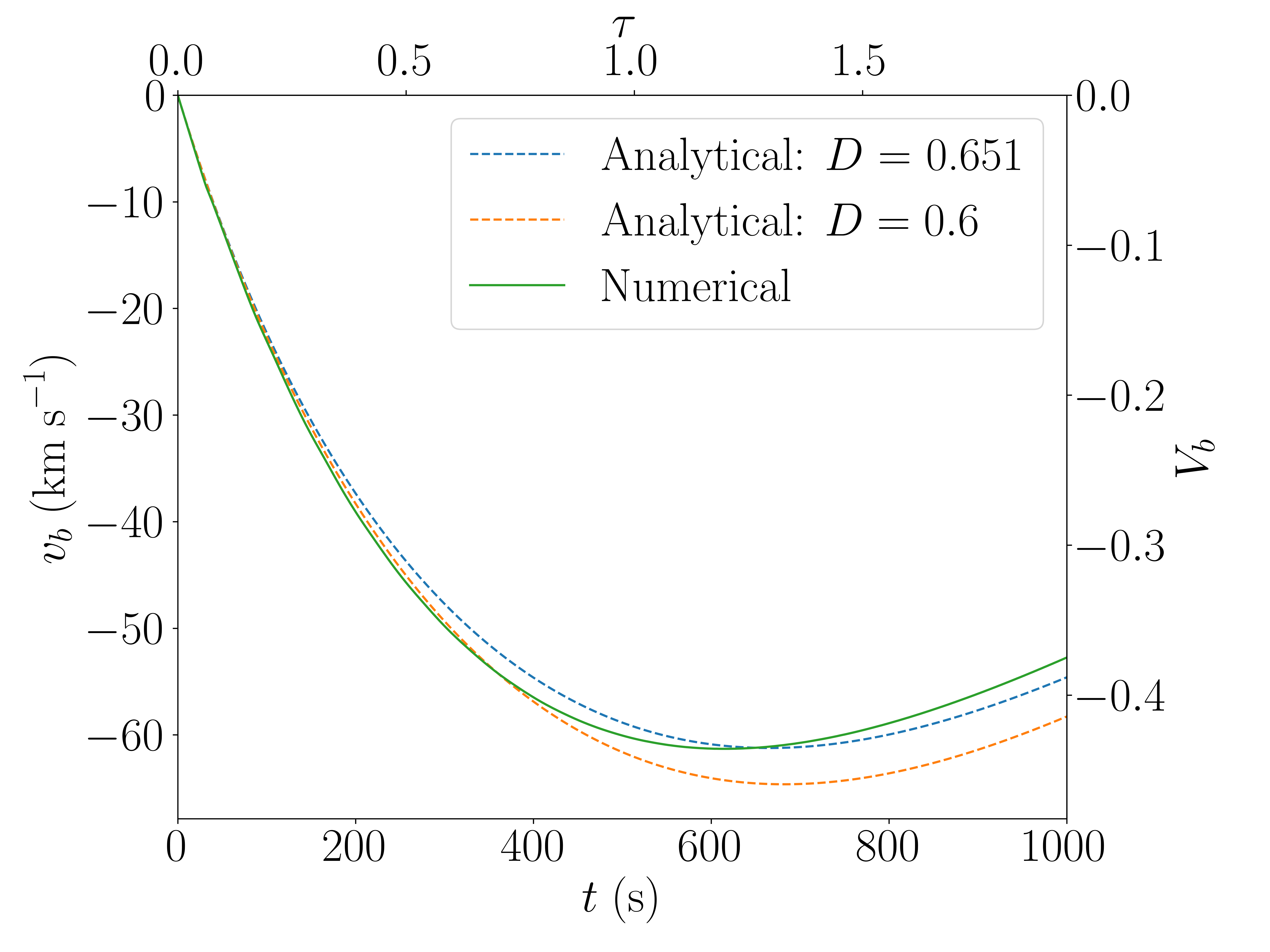}
        \put(-240, 180){\textbf{(b)}} 
    \end{subfigure}
    \begin{subfigure}[t]{0.48\textwidth}
        \centering
        \includegraphics[width=\textwidth]{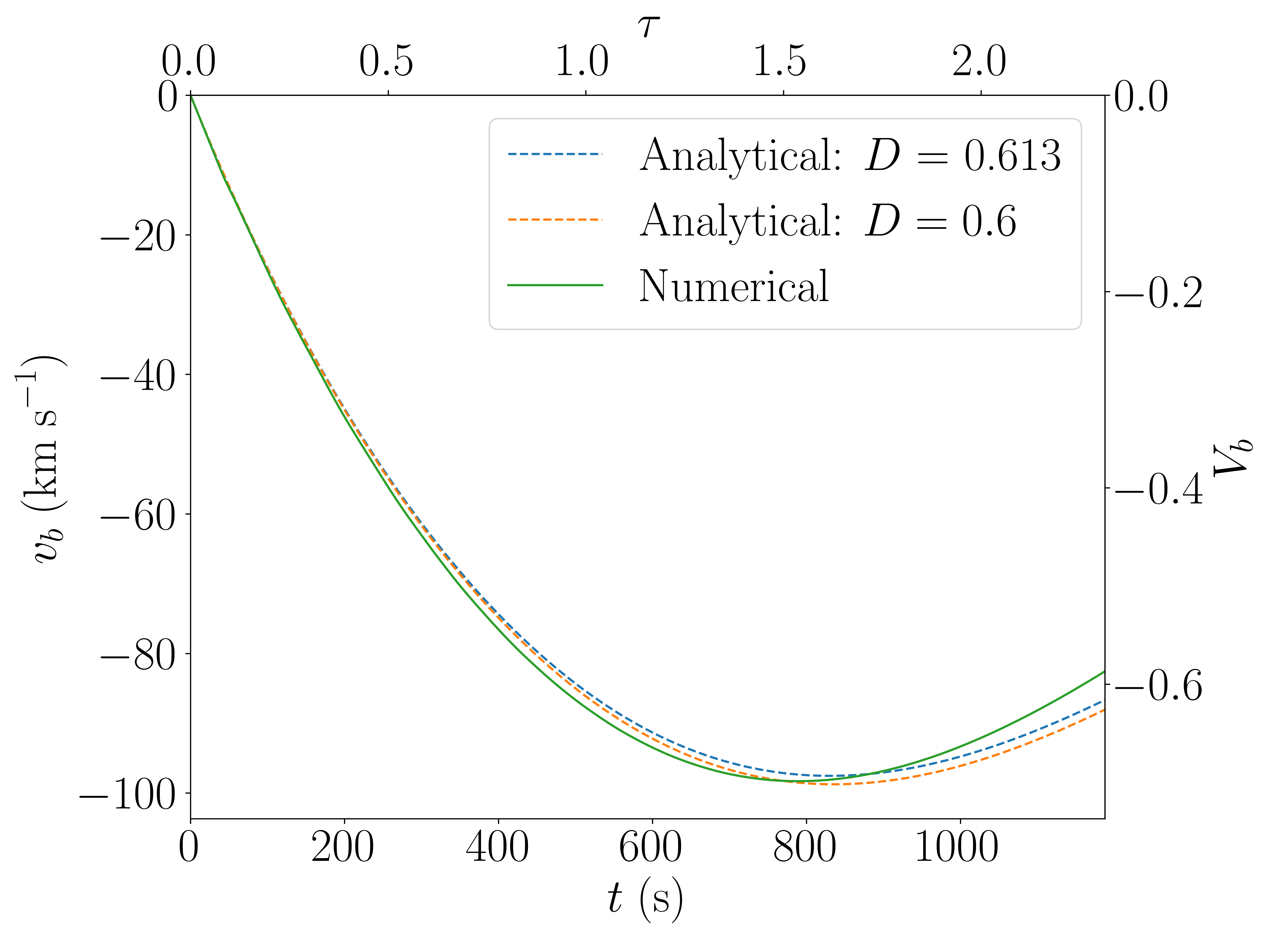}
        \put(-240, 180){\textbf{(c)}}
    \end{subfigure}
    \hfill
    \begin{subfigure}[t]{0.48\textwidth}
        \centering
        \includegraphics[width=\textwidth]{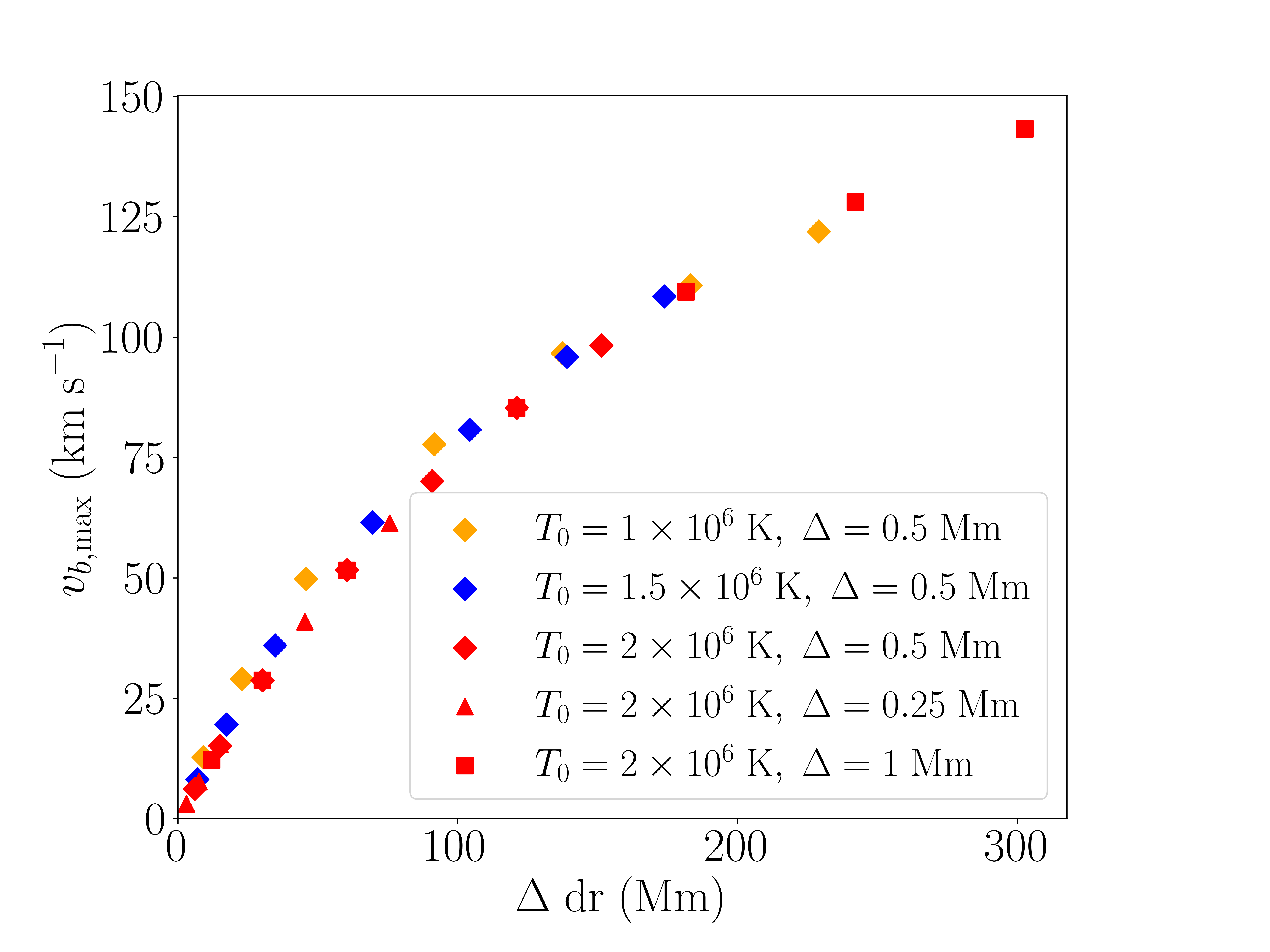}
        \put(-240, 180){\textbf{(d)}} 
    \end{subfigure}
  \end{center}
  \caption{Blob velocity as a function of time for three setups with $T_0=2\times 10^6$~K, $\Delta=0.5$~Mm and (a)~$\rho_{b0}=10^{-10}$~kg~m$^{-3}$, (b) $\rho_{b0}=5\times10^{-10}$~kg~m$^{-3}$ and (c) $\rho_{b0}=10\times10^{-10}$~kg~m$^{-3}$. The green curve comes from the numerical solution of Equations~\ref{eq:mass}--\ref{eq:energy}, whereas the analytical approximation of Equation~\ref{dim_vel} is given by the blue and orange curves for the best fit of $D$ and for $D=0.6$, respectively. The variables $\tau$ and $V_b$ are the dimensionless time and blob speed (see Section~\ref{sect:analytical_stratified}). (d) (Unsigned) maximum blob velocity, $v_{b, {\rm max}}$, versus $\Delta\; dr$ (the meaning of the different symbols is given in the figure legend). }
\label{fig:numerical_velocities}
\end{figure*}

The temporal evolution of the system is as follows: the introduction of a density enhancement at $z=z_0$ sends two sound waves travelling up and down at the (constant) sound speed $C_s=\sqrt{\gamma p_0/\rho_0}$. The blob also starts to fall with a subsonic speed (see Figure~\ref{fig:simulation_results}(a) and the corresponding movie) until it reaches a maximum downward velocity, $v_{b, {\rm max}}$, after which it presents a gentle deceleration. This behaviour of the blob velocity is shown in  Figures~\ref{fig:numerical_velocities}(a), (b), (c). \cite{Oliver2014} found an unexpected association between $v_{b, {\rm max}}$ and the ratio of the blob density to the local ambient density at $t=0$; here we denote this density ratio as $dr$. This association is surprising because atmospheres with different $T_0$ have different scale heights and, although the blob falls through atmospheres with different stratification, it appears that only the initial ambient density at height $z_0$ matters. The plot of $v_{b, {\rm max}}$ versus $dr$ in \cite{Oliver2014} was presented for $\Delta=0.5$~Mm. Changes to this parameter lead to increases or decreases of the  blob mass and, therefore, to larger or smaller $v_{b, {\rm max}}$, respectively, and the already mentioned association is not fulfilled. To correct for this fact, one can plot $v_{b, {\rm max}}$ as a function of $\Delta$~$dr${, as shown in Figure \ref{fig:numerical_velocities}(d),} which results in the points aligning relatively neatly along a curve with little, but non-negligible, spread.

In addition, the falling blob pushes down the gas below it and also pulls down the gas above it. In other words, part of the vertical mass column is set into motion because of the presence of the density enhancement. The lower and upper points of this mass column travel at speed $C_s$, hence at time $t$ they have coordinates $z=z_0\pm C_st$ (see Figure~\ref{fig:simulation_results}(b) and the corresponding movie). The two boundaries of the numerical domain are placed at a large distance from the initial blob position and this prevents the sound waves that bounce off the boundaries from interfering with the blob dynamics \citep[for more details, see][]{Oliver2014}.

\section{The importance of virtual mass}\label{sect:virtual_mass}

Our aim in this paper is to develop a simple analytic model for the motion of coronal rain as presented in the simulations of \citet{Oliver2014} and \citet{Martinez2020}. 
Having seen from the simulations presented in Section \ref{model} that a large region of the corona responds to the motion of the coronal rain, see Figure~\ref{fig:simulation_results}(b), the implication of this is that the ambient mass (i.e. the virtual mass) is evolving and can be significant. Therefore, the key question we need to determine first is: What might be the virtual mass for our problem? There are two key pieces of information from the works of \citet{Oliver2014}, \citet{Martinez2020} and Section \ref{model} that we will use: 1) that if we have a strong magnetic field, the 2D dynamics becomes 1D like, therefore we only look to model 1D dynamics using a point mass, and 2) that {two waves, a compression wave propagating downward and a rarefaction wave propagating upward, are set off as the blob starts to move. The fronts of these waves propagate at the sound speed of the ambient medium. The distance these wave fronts have propagated at a given time determines the amount of the ambient mass (i.e. the virtual mass) that can respond to the motion. The growth of the virtual mass in time can be seen in Figure 1(b) and the accompanying movie, in which the vertical black line separates the unperturbed and perturbed plasma below and above this height, respectively. As time evolves, the height represented by the vertical line moves down at the constant speed $C_s$. This} means that unlike a standard drag model the virtual mass evolves over time. 

This leads to the question: How can we include this dynamically evolving virtual mass into a simple drag-like model? If we treat our coronal rain blob as a mass ($m_b$) falling through a vacuum, then we get this simple equation for the dynamics
\begin{equation}\label{basic_drag}
    m_b\frac{dv_b}{dt}=-m_b g.
\end{equation}
But what happens with the virtual mass ($m_v$) is that it takes momentum from the coronal rain and puts it into the surrounding material. Treating the virtual mass as being able to respond instantly to changes in the momentum of the blob, we need to add two terms to Equation \ref{basic_drag}{:} one to show how change in the speed of the blob increases the speed of the virtual mass, and {another one to represent the} change in the virtual mass, because it is a growing region due to the continued propagation of the sound waves, increasing the momentum of the virtual mass (and reducing the momentum of the blob). This leads to
\begin{equation}
    m_b\frac{dv_b}{dt}=-m_b g-m_v\frac{dv_v}{dt} - v_v\frac{dm_v}{dt},
\end{equation}
where $v_v$ is the velocity of the virtual mass.

Now, we don't know $v_v$ (in fact, as with $v_b$, it will not be a single value but we treat it as such {- this can be seen by the broad velocity distributions in Figure \ref{fig:simulation_results}b going from zero at the compression wave front to a maximum at the blob position}), and as it relates to how the mass in front of and behind the blob moves, it will be some factor (which may not be constant) of the blob velocity {which we denote $D(t)$. This means we represent the average speed of the material in front of and behind the moving blob as $D(t)v_b$}. This gives:
\begin{equation}
  m_b\frac{dv_b}{dt}=-m_b g-m_v\frac{d}{dt}(D(t)v_b) - D(t)v_b\frac{dm_v}{dt},
\end{equation}
or
\begin{equation}\label{final_equation}
    \frac{d}{dt}((m_b+D(t)m_v(t))v_b)=-m_b g,
\end{equation}
where we assume that the coronal rain mass ($m_b$) is constant. Note that Equation \ref{final_equation} does not include any drag term, and can be used to  get a formula for the velocity at a given time. This latter formula is
\begin{equation}
    v_b=-\frac{m_b g}{m_b+D(t)m_v(t)}t.
\end{equation}
So now we look at some specific situations for understanding the evolution of the virtual mass and how this influences the dynamics. 
We note here that we will take $D(t)$ to be a constant $D$ for the rest of this section. We will then be able to check the validity of this assumption when comparing to simulation results.

%Hopefully this will evidence that (as you said all along) this is not drag.\footnote{This is very flattering as it implies that I know more about drag than I actually do. My argument is that 1D simulations give the same results as 2D simulations with a strong enough $B$, hence I do not see how drag can be the answer to the $v_{b, {\rm max}}$ vs. density ratio distribution because I understand that drag requires a flow around an obstacle.}

\subsection{Virtual mass in a constant density background}

Before looking at a stratified atmosphere, it is informative to look at the case of a constant density background ($\rho_0$) with a constant sound speed ($C_s$). 
We have the mass of the blob given by $m_b=SF_b\rho_0h_b$, where $\rho_0$ is the density of the corona where the blob is placed, $F_b$ is the increase factor of the density, {$S$ is the cross-sectional area} and $h_b$ is a measure of the blob {length} (to take this from density to mass). The conversion between $\Delta$ (the half-width of the Gaussian distribution) in \citet{Oliver2014} and this model is $h_b = \sqrt\pi \Delta$ as this means the value of $F_b$ used here directly relates to the density ratio of \citet{Oliver2014} and the simulations of Section \ref{model}. 
For the virtual mass, we know that the mass in causal contact with the moving blob can be given {by
\begin{equation}
    m_v=2S\rho_0C_s t.
\end{equation}
}Based on these arguments we have
\begin{equation}
    v_b=-\frac{g}{1+\frac{2DC_st}{F_bh_b}}t.
\end{equation}

We can quickly see that at early times, the downward velocity will increase at an approximately linear rate which is consistent with free-fall. However, at late times the denominator will approximately scale linearly with time leading to the downward velocity converging to a value of
\begin{equation}\label{max_vel}
    v_b(final)=-\frac{gF_bh_b}{2DC_s}.
\end{equation}
This simple model shows we expect the final velocity to be linearly proportional to the density ratio. However this does not have the earlier peak in velocity with a slow decline afterwards that is seen in Figure \ref{fig:numerical_velocities}.

\subsection{Virtual mass in a stratified atmosphere}\label{sect:analytical_stratified}

Our next step is to develop the model to include a stratified atmosphere with constant sound speed to mimic the simulations presented in Section \ref{model}. This leads to a more complex evolution of the virtual mass. Setting the position where the blob is placed to be $z=0$ we then have
\begin{align}
    m_v=&\rho_0S\int_{-C_st}^{C_st}\exp(-z/H) dz\\=&H\rho_0S\left[\exp\left(\frac{C_s t}{H}\right)- \exp\left(-\frac{C_s t}{H}\right)\right]=2H\rho_0S\sinh\left(\frac{C_s t}{H}\right),\nonumber
\end{align}
with $H$ as the pressure scale height. % which we can rewrite as $H=C_s^2/(g\gamma)$.
Justification for this range of integration is shown in Figure \ref{fig:simulation_results}(b) which shows that for any initial coronal rain mass the region of the ambient corona that is influenced by its motion is restricted to the regions in causal contact as determined by {the propagation of the front of the compression and rarefaction waves which travels at the ambient sound speed}. 
With this new formula for the virtual mass, this gives us
\begin{equation}\label{dim_vel}
    v_b=-\frac{g}{1+\frac{2D(t)H}{F_bh_b}\sinh\left(\frac{C_s t}{H}\right)}t.
\end{equation}
As $t/\sinh(C_s t/H)$ tends to zero as $t$ tends to infinity this implies that $v_b$ will tend to zero at large times. However, for small (but non-zero) $t$ we have a situation that is equivalent to the constant density case so the the downward velocity will initially increase in magnitude with $v_b \propto t$.

The results from this model are overplotted as the orange and blue dashed lines in Figures \ref{fig:numerical_velocities}(a), (b), (c). The blue dashed line is the solution for a value of $D$ optimised to give the best fit to $v_{b}(t)$, the orange line is for a value of $D=0.6$. Both these curves give acceleration of the coronal rain downwards and then deceleration as the virtual mass becomes significant. We can see in Figures \ref{fig:numerical_velocities}(a), (b), (c) that as the mass of the coronal rain is increased, the simulation results become closer to those of the $D=0.6$ curve. {The value of $D=0.6$ implies that the average velocity of the virtual mass is more weighted to the blob mass than would be found for a linear velocity profile. The velocity profiles in Figure \ref{fig:simulation_results} (b) show that such a profile is found in the simulations.}

To make further progress (i.e. to remove the temperature effects from the equation), we can non-dimensionalise Equation \ref{dim_vel}. Firstly we set $\tau=C_s t/H$. This gives (taking $D$ to be constant)
\begin{equation}
    v_b=-\frac{gH/C_s}{1+\frac{2DH}{F_bh_b}\sinh\left(\tau\right)}\tau.
\end{equation}
From this equation we see we have the freefall velocity after a sound crossing time ($gH/C_s$) and the ratio of the blob mass to the mass in a pressure scale height ($M=F_bh_b/H$) as fundamental quantities of the system. This leads to the non-dimensionalised equation
\begin{equation}\label{non-dim_eqn}
    V_b=-\frac{1}{1+\frac{2D}{M}\sinh\left(\tau\right)}\tau,
\end{equation}
with $V_b=v_bC_s/(gH)$. 

\begin{figure}
    \centering
    \includegraphics[width=9.5cm]{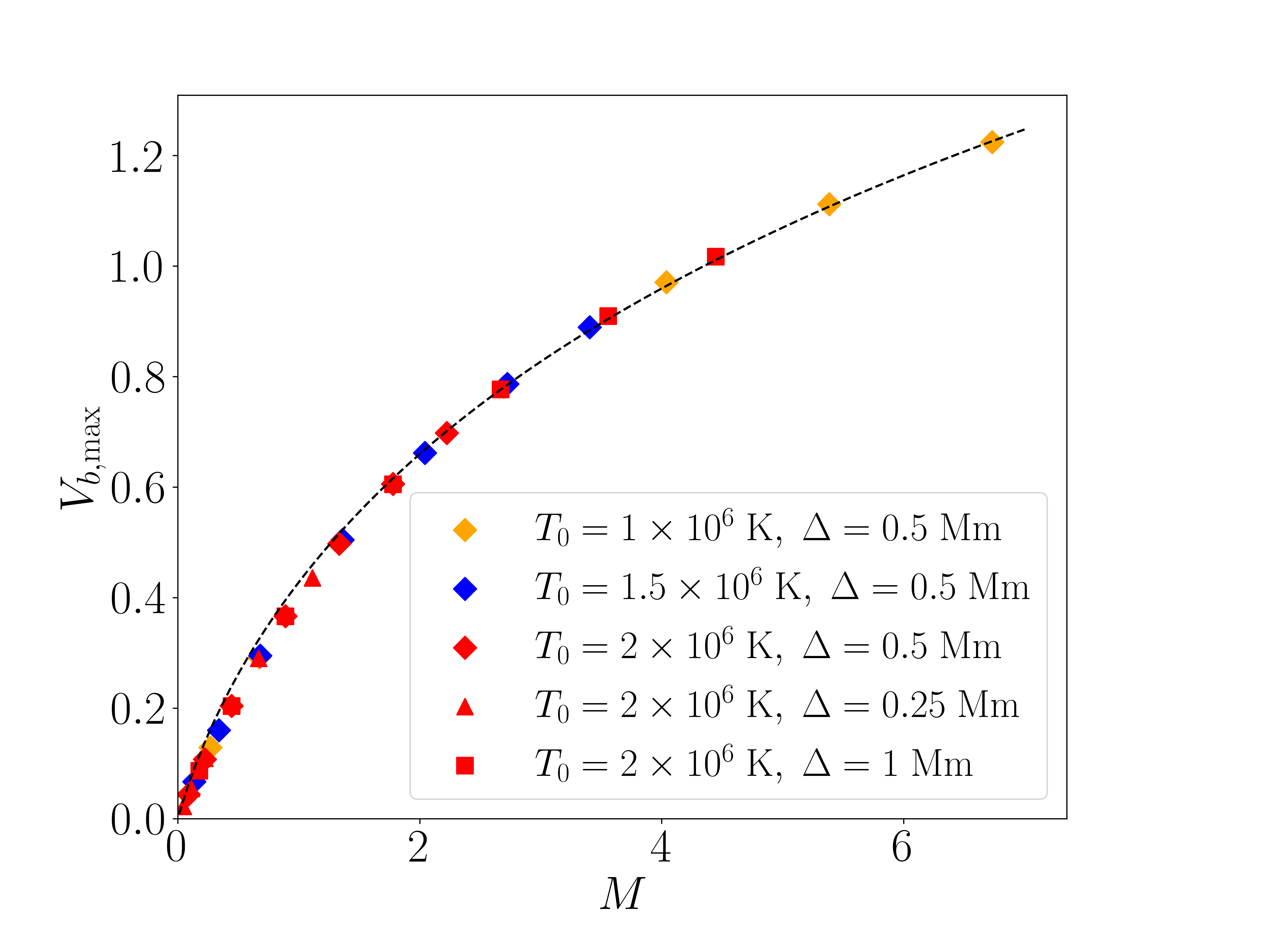}
    \caption{Plot of (unsigned) maximum $V_b$ against $M$ for the simulation results (the meaning of the different symbols is given in the figure legend). The dashed line is the predicted value from Equation \ref{non-dim_eqn} with $D=0.6$.} 
    \label{fig:3}
\end{figure}

The next step is to determine the relation between $M$ and the maximum value of $V_b$ ($V_{b, {\rm max}}$). By taking the derivative of Equation \ref{non-dim_eqn} with respect to $\tau$, we find
\begin{equation}\label{eq:tau_max}
    M=2D\cosh(\tau_{\rm max})(\tau_{\rm max} - \tanh(\tau_{\rm max})),
\end{equation}
with $\tau_{\rm max}$ the value of $\tau$ associated with $V_{b, {\rm max}}$.
Using this to remove $M$ from Equation \ref{non-dim_eqn} gives
\begin{equation}\label{eq:Vb_max}
    V_{b, {\rm max}}=\tau_{\rm max} - \tanh(\tau_{\rm max}).
\end{equation}
Though this does not give a direct analytic formula for dependence of $V_{b, {\rm max}}$ on $M$, it does allow them to be directly connected.

The simulation results are plotted using this new normalisation in Figure \ref{fig:3}. This shows the various different models collapsed onto the same curve, confirming that the non-dimensionalisation correctly captures the underlying physics of the problem. The dashed line is the analytic prediction stated above using $D=0.6$. We can see that at small $M$ the model slightly over-predicts the value of $V_{b, {\rm max}}$. However, above $M\approx 1$, the simulation results and the model align to good degree. It should also be noted that using the normalisation in this figure, the small spread that can be seen in the simulation results presented in Figure \ref{fig:numerical_velocities}(d) has been completely removed.

\section{Summary and discussion} \label{sec:summary}

In this paper, we have presented a simple model for the dynamics of coronal rain that provides an explanation of much of the behaviour found in the simulations of rain evolution by \citet{Oliver2014}. The key process is that the corona responds over a large distance (determined by the propagation of sound waves) to the motion of the coronal rain to effectively add mass to the system which in turn reduces its speed.
A particular key result is that the scaling of the maximum downflow speed as a function of the density ratio to the power 0.64 given by a fit to the data in \citet{Martinez2020} here is explained by an analytic relation that appears to have a similar trend to this power law. 

As we have been able to show that the model for virtual mass effectively decelerating the coronal rain in the model of \citet{Oliver2014}, this leads to the question: what role if any does a drag-like (by this we mean a term $\propto v_b^2$) play? In fact we can see in Figure \ref{fig:3} that for small $M$ values, the value of $V_{b, {\rm max}}$ from the simulations is smaller than we predict from our model. For these small-mass cases, it is likely that pressure gradients acting to give an upward force are created by the downward motion of the coronal rain and these can act as a drag. However, for larger rain mass ({i.e. $M\gtrapprox 1.5$ which are} more consistent with those of the solar atmosphere) this effect becomes small.

Finally, an important question to ask is: can we reasonably assume that $D(t)$ will be constant in time, and will it have the same value for the regions above and below the coronal rain. We can see from Figure \ref{fig:numerical_velocities} that for large rain mass the model provides a good representation of the velocity evolution over long time ({i.e. $M\gtrapprox 1.5$ the lower bound of which corresponds to Figure~\ref{fig:numerical_velocities}(b)}), meaning that changes in the value should be small. However, at very large ({i.e. $M\gtrapprox 6$}) rain masses the downflow velocity may become trans- or super-sonic which will change the way in which the rain motion is transmitted into the surrounding corona. This will likely result in the value of $D$ evolving over time. {Relevant to this, some observations have implied that there is strong compression \citep{Antolin2023} or shocks \citep{schad2016} preceding coronal rain blobs. Compression in front of the rain as it moves is a fundamental aspect of our model. Therefore, observations may create a pathway to calculate the value of $D$ for real dynamics. In addition to this, shock formation is something that our model implies as a possibility and an important further step will {be calculating} the relevant $M$ values for the observations to see if we are predicting transonic flow.}

{As a final comment, simplified drag models are widely used in understanding and modelling dynamics in a broad range of situations including CME propagation \citep[e.g.][]{Vrsnak2010} and the dynamics of cool, dense clouds of gas in fast outflows found in the Intracluster medium \citep[e.g.][]{Gronke2018}. 
However, our model with evolving virtual mass is more relevant for situations with strong magnetic field, as quantified by having sub-Alfv\'{e}nic flow speeds in a low-beta plasma environment, This is a fundamental difference from the examples listed above where weaker magnetic fields mean aerodynamic drag is the dominant process influencing the temporal evolution. As such, the ideas that underpin our model are likely to be applicable in situations where motion is confined to flow along the magnetic field; an example of this may be accretion to the poles of highly magnetised neutron stars \citep[e.g][]{Davidson1973}, presenting some further avenues for this work.}

\begin{acknowledgements}
      AH is supported by STFC Research Grant No. ST/R000891/1 and ST/V000659/1. {DM acknowledges support from the Spanish Ministry of Science and Innovation through the grant CEX2019-000920-S of the Severo Ochoa Program.} This publication is part of the R+D+i project PID2023-147708NB-I00, financed by MICIU/AEI/10.13039/501100011033/ and FEDER, EU. AH and RO would like to acknowledge the discussions with members of ISSI Team 457 ``The Role of Partial Ionization in the Formation, Dynamics and Stability of Solar Prominences'' and ISSI Team 545 ``Observe Local Think Global: What Solar Observations can teach us about Multiphase Plasmas across Astrophysical Scales'', which have helped improve the ideas in this manuscript.  {We would also like to thank the anonymous referee for their useful comments.} The simulation data presented in Section \ref{model} is based on the work of \citet{Oliver2014} and will be made available on reasonable request. There are no other data produced for this paper.
\end{acknowledgements}

\bibliographystyle{aa}
\bibliography{blob}

\end{document}